\documentclass[conference]{IEEEtran}
\IEEEoverridecommandlockouts
\usepackage{cite}
\usepackage{tikz}
\usetikzlibrary{quantikz2}
\usepackage{amsmath,amssymb,amsfonts}
\usepackage{algorithmic}
\usepackage{graphicx}
\usepackage{subcaption}
\usepackage{textcomp}
\usepackage[table]{xcolor}
\usepackage{braket}
\usepackage{array}
\usepackage{makecell}
\usepackage{graphicx}
\usepackage{booktabs}
\usepackage{listings}
\usepackage{xparse}
\usepackage{comment}
\usepackage{balance}
\usepackage[table]{xcolor}
\usepackage[colorlinks,urlcolor=blue,linkcolor=blue,citecolor=blue]{hyperref}
\usepackage{hyphenat}

\newcommand{\orcidnew}[1]{\href{https://orcid.org/#1}{\includegraphics[width=8pt]{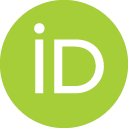}}}

\newcommand*{\email}[1]
{\normalsize\texttt{\href{mailto:#1}{#1}}\par}

\newcolumntype{s}{>{\columncolor{gray!30} \bf} c}
\newcolumntype{b}{>{\bf} c}
\def\BibTeX{{\rm B\kern-.05em{\sc i\kern-.025em b}\kern-.08em
    T\kern-.1667em\lower.7ex\hbox{E}\kern-.125emX}}
\makeatletter

\begin{document}

\title{QML for Quantum Sensing under Measurement-Induced Information Loss}

\author{
    \IEEEauthorblockN{%
         Sounak Bhowmik \orcidnew{0009-0002-0931-2644} and
         Himanshu Thapliyal \orcidnew{0000-0001-9157-4517}
                    }
 
    \IEEEauthorblockA{Department of Electrical and Computer Engineering\\
        Southern Methodist University, Dallas, TX, USA}
    \email{sbhowmik@smu.edu},
    \email{hthapliyal@smu.edu}            
}

\maketitle
\begin{abstract}
Nitrogen-vacancy (NV) centers in diamond can serve as highly sensitive solid-state quantum sensors for high-sensitivity magnetometry. However, in the noisy intermediate-scale quantum (NISQ) era, extracting reliable information from noisy, finite-shot, and measurement-limited sensing data remains a considerable challenge. Whereas, quantum machine learning (QML) offers a potential path to improve parameter estimation by learning nonlinear relationships between quantum-sensing data and the underlying physical signal. In this work, we investigate the role of QML in magnetic-field estimation within an NV center-inspired magnetometry setting. We formulated magnetic field sensing as a supervised regression task. We compared the performance of several classical machine learning models trained on measurement-based classical data with that of quantum kernel-based models trained on pre-measurement coherent quantum states. Our objective is to isolate the impact of measurement-induced information loss and therefore provide a theoretical upper bound on the sensing performance. The upper bound is achievable only when coherent quantum information is directly available to the learning model. Our results show that QML-based sensing performance improves significantly with coherent quantum-state information, and not much with changes in model complexity or learning paradigm. This observation underscores the importance of learning pipelines that tightly integrate quantum sensors and QML models to enhance magnetic field sensing under realistic constraints.
\end{abstract}


\section{Introduction}
\label{sec:introduction}

Quantum sensing~\cite{degen2017quantum, zhang2021distributed} has emerged as one of the most promising near-term applications of quantum technology, with the potential to improve precise measurements beyond the limits of many classical sensing platforms. Among the leading solid-state sensing systems, negatively charged nitrogen-vacancy (NV) centers in diamond have attracted significant attention. NV centers as sensors offer a combination of atomic-scale spatial resolution, optical initialization and readout, microwave spin control, and room-temperature operation~\cite{hong2013nanoscale, schirhagl2014nitrogen}. These properties make the NV center a powerful platform for measuring magnetic fields, electric fields, temperature, strain, and other local physical quantities with high sensitivity.

\begin{figure}[b!]
    \centering
    \includegraphics[width=0.9\linewidth]{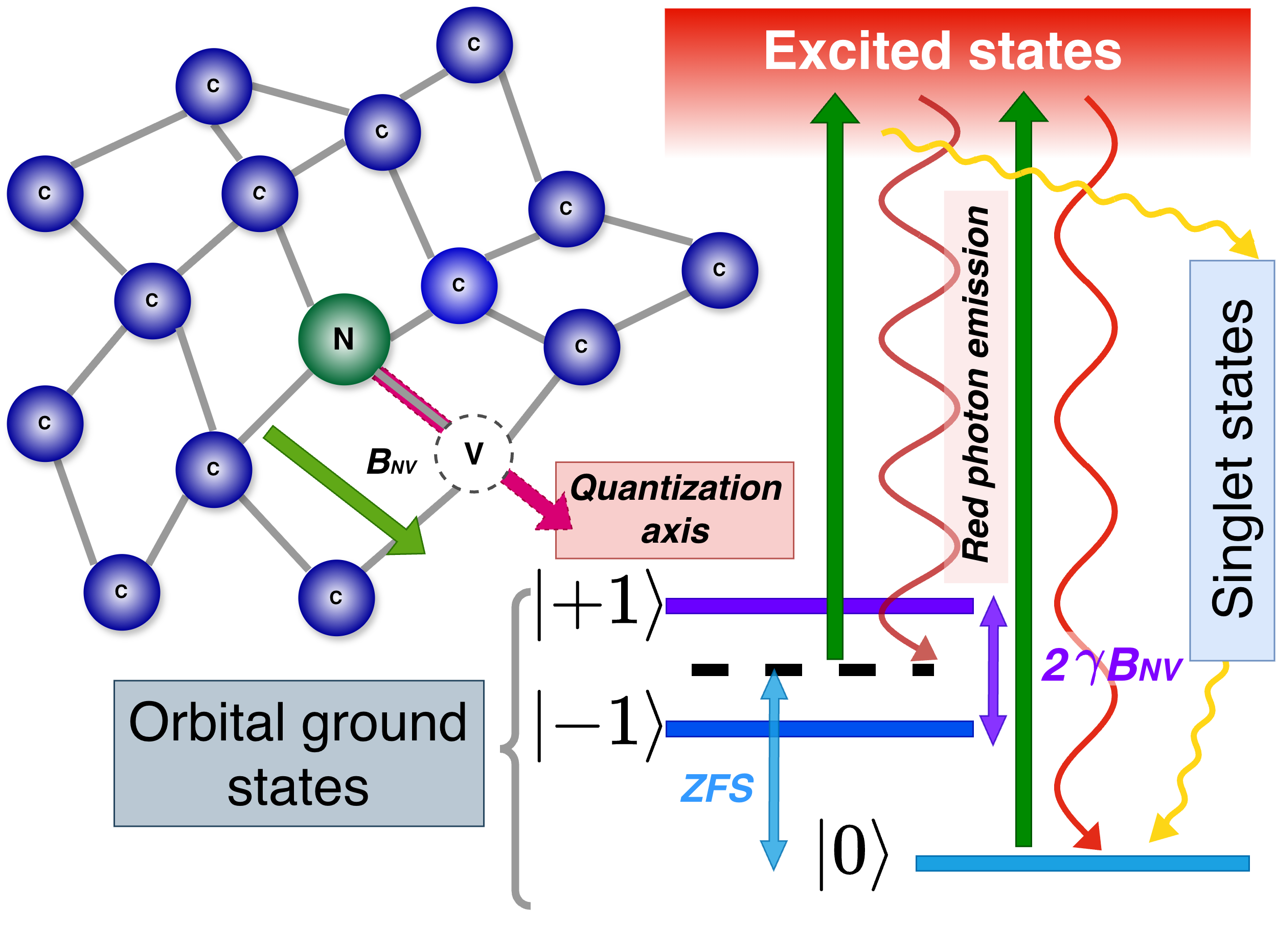}
    \caption{Level diagram of a nitrogen-vacancy center in diamond, where a substitutional nitrogen atom and an adjacent vacancy form a spin defect used for nanoscale magnetometry. \footnotesize{\\Without any external magnetic field, the energies of the orbital ground states, $\in\{\ket{0}, \ket{\pm1}\}$, split, called zero field splitting (ZFS). In the presence of an external magnetic field, $B_{NV}$ along the quantization axis of the crystal, the degenerate ground states, $\ket{+1}$ and $\ket{-1}$, split up, which is referred to as Zeeman splitting. Once optically excited, the $\ket{\pm1}$ states decay through metastable singlet states to the ground state, $\ket{0}$, leading to optical initialization and spin-dependent fluorescence to be read optically to estimate the unknown magnetic field.}} 
    \label{fig: nvcenter}
\end{figure}

In NV center magnetometry, the physical signal is encoded into the quantum state of the sensor through coherent spin evolution. 
For example, in Ramsey interferometry, an external magnetic field can produce a phase shift in the NV spin state, which is subsequently estimated from measurement outcomes~\cite{rondin2014magnetometry}. Although the underlying quantum state contains rich information about the magnetic field, practical experiments cannot directly access the full coherent state. 
Instead, they provide finite measurement statistics obtained after projective or optical readout. 
This distinction is critical. Noise, decoherence, shot noise, and limited sampling can remove or obscure information before it becomes available to a learning algorithm. 
As a result, the performance of any downstream inference method is constrained not only by the model architecture but also by the information it can preserve in the measured data.

Recent advances in quantum machine learning (QML) have motivated new approaches for signal processing, parameter estimation, and data-driven inference in quantum sensing systems~\cite{schuld2019quantum, rawat2026emerging}. 
QML models are often motivated by their ability to represent data in high-dimensional quantum feature spaces and therefore exploit coherent quantum dynamics for learning. 
However, in realistic sensing workflows, it remains unclear whether QML provides an advantage with only post-measurement classical data available. 
Therefore, an important question arises: does the improvement in sensing performance primarily arise from the expressive power of the learning model or from access to coherent quantum information before measurement?

\begin{figure*}[htbp!]
    \centering
    \includegraphics[width=\linewidth]{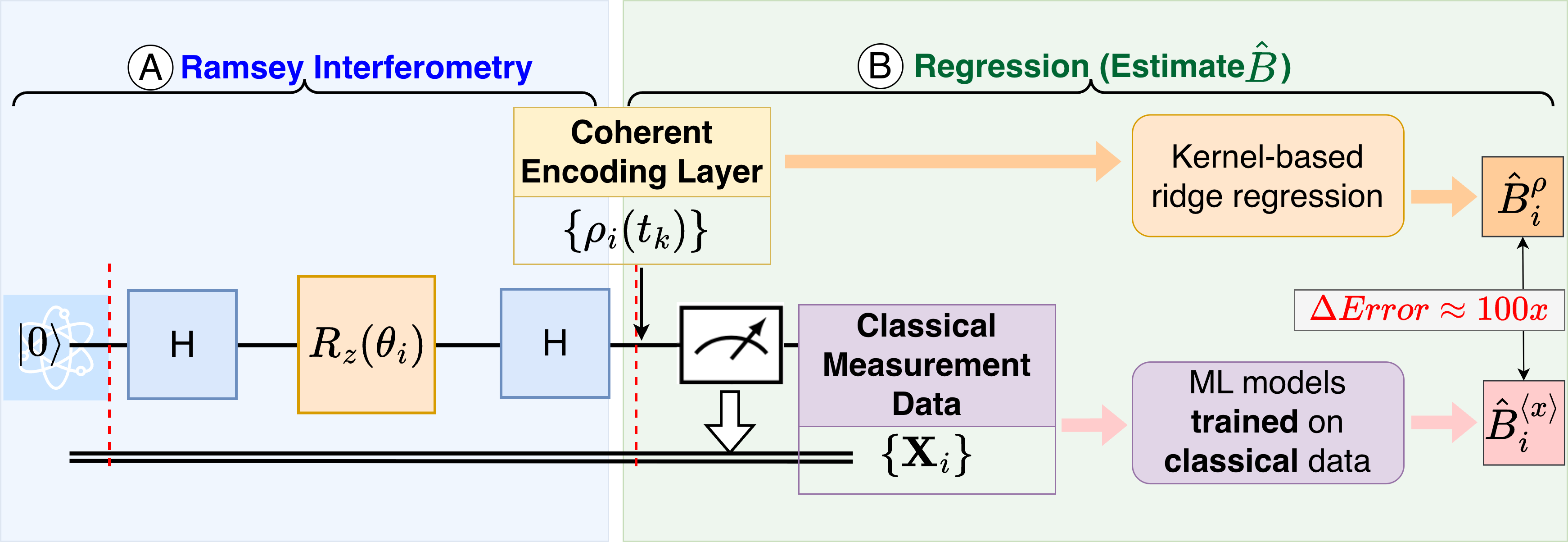}
    \caption{Magnetic field estimation using NV center-inspired quantum sensing simulation. \footnotesize{(A) Ramsey Interferometry simulation: $\theta_i=\gamma B_it_k$; $\rho_i(t_k)$ is the corresponding density matrix, after the system evolves under a magnetic field $B_i$ for time $t_k$; $\mathbf{x_i}$ is the corresponding expectation value of the observable. (B) Regression problem to estimate the unknown magnetic field: $\hat{B}^{\rho}$ and $\hat{B}^{\braket{x}}$ are the estimated magnetic field values based on coherent-state and classical measurement statistics, respectively.}}
    \label{fig: pipeline}
\end{figure*}


In this work, we investigate this question through a controlled sensing experiment inspired by NV center magnetometry. 
We formulate magnetic-field estimation as a supervised regression task and evaluate various classical machine learning models using measurement statistics from Ramsey interferometry experiments.
Afterward, we compare these results with a quantum kernel ridge regression model, trained directly on pre-measurement coherent quantum states represented as density matrices. 
This quantum-kernel setting provides a theoretical upper bound on the information available before measurement-induced loss occurs.

Our results aim to demonstrate the bottleneck imposed by measurement-induced information loss on the learning-based quantum sensing task. 
We observed that the quality and completeness of the available readout data limit the classical models trained on the measured statistics. 
Even when the underlying sensing dynamics contain stronger field-dependent structure, these models deviate far from the theoretical upper bound. 
In contrast, when we trained the quantum kernel models operating on coherent density matrices, the estimation performance improved substantially.
This observation underscores the impact of measurement-induced information. It suggests that QML's success in quantum sensing depends largely on preserving, accessing, and effectively encoding the coherent quantum data before readout.

The main contributions of this work are threefold:
\begin{enumerate}
    \item We formulate an NV center-inspired Ramsey magnetometry experiment as a regression benchmark for studying learning under realistic measurement constraints.
    \item We quantify the impact of measurement-induced information loss by comparing the performance of the classical models trained on finite measurement statistics with quantum kernel models trained on coherent quantum state representations.
    \item We establish a theoretical upper bound on sensing performance using quantum kernel ridge regression over pre-measurement density matrices generated during the sensing evolution.
\end{enumerate}

The rest of this paper is organized as follows. Section~\ref{sec: background} introduces NV center-inspired magnetometry, the sensing protocol, and the measurement strategy. Section~\ref{sec: experiment} describes the experimental setup, simulation parameters, and noise model that have been used in this experiment. Section~\ref{sec: learning} depicts how different models learn from the classical and quantum data, highlighting their differences in terms of learning strategy. Section~\ref{sec: results} presents the performance analysis and discusses the effect of measurement-induced information loss. Finally, section~\ref{sec: conclusion} concludes the paper and outlines future research directions.

\section{Background}\label{sec: background}
\subsection{NV centers in Diamond}
Nitrogen-vacancy (NV) centers in diamond are atomic-scale point defects formed by a nitrogen atom, adjacent to a vacant carbon lattice site. 
In its negatively charged state, the NV center has an electronic spin-triplet ground state, $m_s = {0, +1, -1}$, making it a robust solid-state quantum sensor. 
As shown in the figure~\ref{fig: nvcenter}, the nitrogen vacancy (NV) axis defines the spin system's quantization axis. 
Even in the absence of an external magnetic field, the ground-state energy levels are split by a phenomenon called zero-field splitting (ZFS)~\cite{rondin2014magnetometry, hong2013nanoscale}, separating the $\ket{0}$ state from the degenerate $\ket{+1}$ and $\ket{-1}$ states by approximately 2.87 GHz. 
When a magnetic field component $B_{NV}$ is projected along the NV axis, the $\ket{\pm1}$ states undergo Zeeman splitting~\cite{zeeman_split}, producing an energy separation proportional to $2\gamma B_{NV}$, where $\gamma$ is the gyromagnetic ratio of electronic spin. 
This magnetic-field-dependent splitting is the basis of NV center magnetometry.

Figure~\ref{fig: nvcenter} further illustrates the optical spin-readout mechanism. 
Under green laser excitation, the NV center emits red fluorescence. However, the fluorescence intensity depends on the spin state as the $\ket{\pm1}$ states have a higher probability of decaying through intermediate singlet states before returning to $\ket{0}$. 
This spin-dependent fluorescence enables optical initialization and readout, while microwave excitation drives transitions between $\ket{0}$ and $\ket{\pm1}$. 
By monitoring shifts in the optically detected magnetic resonance spectrum, local magnetic fields can be measured with nanoscale spatial resolution under ambient conditions.

NV magnetometry experiments are usually restricted to two-level systems, formed by $m_s=\{0,-1\}$.
This setup is a suitable qubit representation for Ramsey interferometry [Fig.~\ref{fig: pipeline}(A)]~\cite{degen2017quantum, rondin2014magnetometry}. 

\subsection{NV Inspired Quantum Sensing Simulation}
To isolate only the field-dependent dynamics of the NV center, we consider only the Zeeman interaction contributing to the effective Hamiltonian~\cite{nielsen2010quantum, schirhagl2014nitrogen} of the system under observation: $\mathcal{H} = \hbar \gamma BS_z$. Where $B$ is the unknown magnetic field, our parameter of interest. 
Time evolution of this Hamiltonian will produce a reliable simulation of a physical NV center system: 
\begin{equation}\label{eqn: hamiltonian}
    U(t)= e^{-i\mathcal{H}t}=e^{-i\hbar\gamma BtS_z}
\end{equation}

For $S_z=\frac{1}{2}\sigma_z$, $\mathcal{H}=\frac{\hbar}{2}Bt\sigma_z$, the unitary evolution can be described as equation~\ref{eqn: hamiltonian_evolvution_circuit}:

\begin{equation}\label{eqn: hamiltonian_evolvution_circuit}
U(t)=e^{-i{\hbar\gamma Bt \sigma_z/}{2}} =  R_Z(\phi)
\end{equation}

for $\phi=\gamma Bt$, considering $\hbar=1$. Therefore, our quantum circuit to simulate the Hamiltonian of a two-level NV center model consists of a rotation gate around the Z-axis, with an angle that depends on the magnetic field and time.

However, initially, the basis state gets transcended to a highly sensitive superposition state, with the application of a $\frac{\pi}{2}$-pulse, essentially a Hadamard gate: 

\begin{equation}\label{eqn: prepare_sensor}
\ket{\psi_0}={H}\ket{0}=\frac{\ket{0}+\ket{1}}{\sqrt{2}}
\end{equation}

Afterward, we evolve this initial sensing state $\ket{\psi}$ under the influence of $U(t)$: 

\begin{equation}\label{eqn: prepare_sensor}
\ket{\psi_1}=e^{-i\gamma Bt/{2}}\ket{\psi_0}
\end{equation}

Finally, we apply a second Hadamard gate to bring $\psi_1$ back to the computational basis~[Fig.~\ref{fig: pipeline}(A)]. 

Then we measure the system and process the data to estimate the parameter of interest, the magnetic field, $B$. However, before measurement, the magnetic field is encoded in the off-diagonal elements of the density matrix; measurement in the Z-basis irreversibly discards the phase information, leading to irreversible data loss. 
\begin{equation}
\rho(t_k)=\frac{1}{2}
\begin{pmatrix}
    1&e^{-i\phi}\\
    e^{i\phi}&1
\end{pmatrix}
\end{equation}

Where $\phi=\gamma Bt_k$, and $t_k$ is the time of evolution.

\section{Experimental Setup} \label{sec: experiment}
To examine the true quantum advantage offered by QML, we designed a controlled sensing experiment inspired by NV center magnetometry. 
Figure~\ref{fig: pipeline} shows the experimental pipeline used to estimate the magnetic field from the NV center simulation. 
We begin by preparing the basis states, transforming into a more sensitive superposition state, and simulating their unitary time evolution under an unknown magnetic field [refer to equation~\ref{eqn: hamiltonian}]. 
The resulting data are then collected and used to train the following regression models [refer to Fig.~\ref{fig: pipeline}(B)]:

\begin{itemize}
    \item Classical machine learning models on the expectation values ($\{\braket{Z}_{t_1}, \braket{Z}_{t_2}, \dots, \braket{Z}_{t_M}\}$), and
    \item Quantum fidelity-kernel ridge regression model on coherent quantum states ($\mathcal{R}_i=\left\{\rho_i(t_1), \rho_i(t_2), \dots, \rho_i(t_M)\right\}$).
\end{itemize}

To better capture the physical behavior of the NV center, we incorporated a realistic noise model~\cite{degen2017quantum,rondin2014magnetometry}, including a phase-damping channel with characteristic time $T_2=100~\mu s$. 
Measurement noise is modeled as a 10\% classical bit-flip readout error, and shot noise is introduced through binomial sampling of the measurement outcomes. 
The system evolves at time points $t_k\in\{5, 10, 20, 40, 60\}~\mu s$. The noise model and the parameters describing the sensing dynamics are kept fixed throughout all experiments.
Table~\ref{tab:params} shows a list of the parameters used in this experiment.
\begin{table}[htb!]
\centering
\caption{Simulation and learning parameters used in the NV center-inspired sensing benchmark.\footnotesize{}}
\label{tab:params}
\begin{tabular}{|l|c|}
\toprule
\hline
\textbf{Parameter} & \textbf{Value} \\
\hline
\textit{Magnetic field range} & $B \in [0,2]~\mu T$ \\
\textit{Training samples} & 800 \\
\textit{Test samples} & 100 \\
\textit{Evolution times} & $\{5,10,20,40,60\}~\mu s$ \\
\textit{Random seeds} & $3$\\
\textit{Dephasing time} & $T_2 = 100~\mu s$ \\
\textit{Readout error} & 10\% bit-flip noise \\
\textit{Shot noise} & Binomial sampling \\
\textit{Evaluation metric} & RMSE on $B$ \\
\textit{Classical baselines} & LRR, RBF-KRR, MLP \\
\textit{Quantum model} & Fidelity-kernel ridge regression \\
\hline
\bottomrule
\end{tabular}
\footnotesize{\\ --- \\ RMSE: Root mean squared error, LRR: Linear ridge regression, RBF: Radial basis function, KRR: Kernel ridge regression, MLP: Multilayer perceptron}
\end{table}

While recording the dataset, we stored the density matrix prior to measurement, and later used it as the quantum data to train the QML models. In contrast, the post-measurement expectation values are treated as classical data in classical regression models. 

In the next section, we shed more light on this distinction while building the learning framework for classical and quantum machine learning models.

\section{Learning From Classical and Quantum Data}
\label{sec: learning}
The central objective of this study is to separate the effect of model expressivity from the effect of quantum data accessibility. 
In this experiment, we constructed two learning pipelines that use the same underlying NV centered-inspired sensing dynamics but differ in the information available to the learner. 
The classical pipeline receives only measurement-derived features after projective readout. 
In contrast, the quantum pipeline operates on the pre-measurement density matrices generated during the coherent sensing evolution. 
This distinction allows us to evaluate how much information is lost during measurement and how this loss affects magnetic-field estimation.

\subsection{Classical Learning From Measurement Records}

In the classical learning pipeline, each training instance is generated by sampling a magnetic field value $B_i$ between $0-2\mu T$ and running the Ramsey sensing protocol through multiple evolution times $t_k$. 
After each evolution, the resultant state is measured and converted into a finite-shot estimate of an observable expectation value. 
Therefore, the classical learner does not access the quantum state directly; rather, it only receives a feature vector constructed from measurement statistics.

We consider two classical feature representations. The first is a $Z$-only feature map, where the input is given by
\begin{equation}
    \mathbf{x}^{(Z)}_i =
    \left[
    \braket{Z}_{i,t_1},
    \braket{Z}_{i,t_2},
    \dots,
    \braket{Z}_{i,t_M}
    \right].
\end{equation}
This setting represents a restricted measurement protocol in which only computational-basis readout is available. 
Since the magnetic field is encoded as a phase during the Ramsey evolution, this feature map loses information about it during the measurement protocol.

The second representation is an $XYZ$ feature map, where measurements are performed in three Pauli bases:
\begin{equation}
\label{eq: classical_measuremenet}
\begin{array}{l}
    \mathbf{x}^{(XYZ)}_i =
    [    \{\braket{X}_{i,t_1}, \braket{Y}_{i,t_1}, \braket{Z}_{i,t_1}\},
    \dots,\\
     \quad \quad \quad \quad \quad \{\braket{X}_{i,t_M}, \braket{Y}_{i,t_M}, \braket{Z}_{i,t_M}\}]
\end{array}
\end{equation}

This representation provides a slightly richer classical description of the resultant state because it partially captures the coherence through measurements in complementary bases. 
However, limitations regarding finite-shot sampling, readout error, and the need to estimate quantum-state information from classical measurement outcomes persist.

For the linear classical baseline model, we use ridge regression and solve
\begin{equation}
    \min_{\mathbf{w}}
    \sum_{i=1}^{N}
    \left(
    B_i - \mathbf{w}^{T}\mathbf{x}_i
    \right)^2
    + \lambda \|\mathbf{w}\|^2 ,
    \label{eqn:ridge_reg}
\end{equation}
where $\mathbf{x}_i$ denotes either the $Z$-only or $XYZ$ classical feature vector, $B_i$ is the target magnetic field, and $\lambda$ is the regularization strength. 
We also evaluate nonlinear classical baselines, including RBF-kernel ridge regression and a multilayer perceptron, to separate measurement-induced information loss from limitations arising from the limited linear model capacity.

\subsection{Quantum Learning From Pre-Measurement Quantum States}

The quantum machine learning pipeline uses the same physical sensing process but assumes access to the full density matrices before final measurement. 
For each sampled magnetic field $B_i$, the Ramsey protocol generates a trajectory of quantum states across the selected evolution times ($t_k$):
\begin{equation}
    \mathcal{R}_i =
    \left\{
    \rho_i(t_1),
    \rho_i(t_2),
    \dots,
    \rho_i(t_M)
    \right\}.
\end{equation}
This representation preserves the coherent phase information encoded during sensing within the off-diagonal elements of the density matrix.

\begin{figure*}[htbp!]
    \centering
    \includegraphics[width=0.7\linewidth]{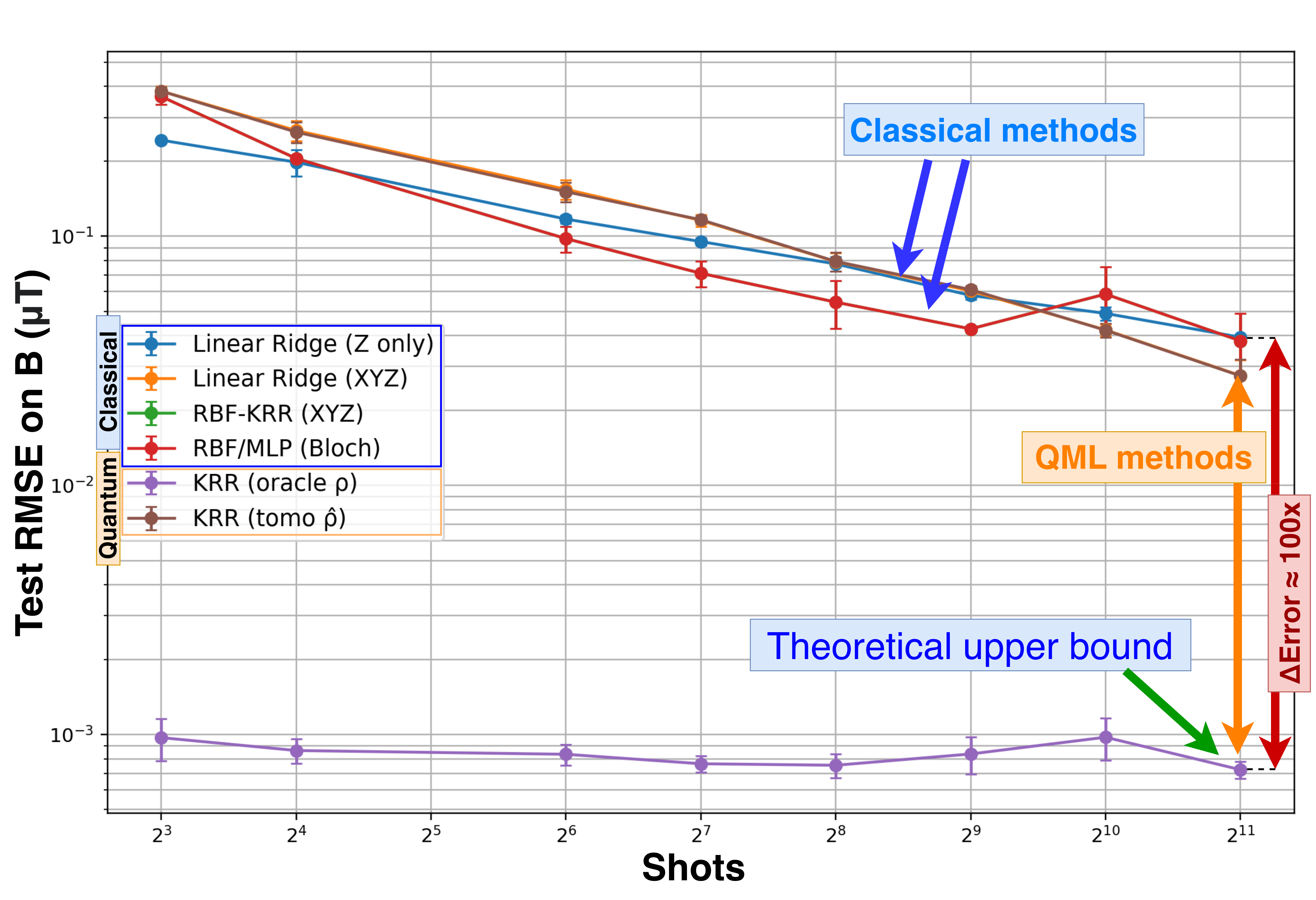}
    \caption{NV center-inspired magnetometry regression (mean $\pm$ std over seeds): \footnotesize{Depicts the theoretical upper bound, and its performance gap with the models learning from the measurement data. The estimation error in the latter is $100$ times more. Sensing pipelines that are more closely integrated with the coherent quantum states, or a more sophisticated measurement strategy, can help to close this gap and provide more accurate results.}}
    \label{fig: results}
\end{figure*}


To learn from these quantum-state trajectories, we use kernel ridge regression with a fidelity-based quantum kernel. For two sensing trajectories $\mathcal{R}_i$ and $\mathcal{R}_j$, the kernel is defined as

\begin{equation}
    k(\mathcal{R}_i,\mathcal{R}_j)
    =
    \frac{1}{M}
    \sum_{k=1}^{M}
    F\left(
    \rho_i(t_k),\rho_j(t_k)
    \right)^2 ,
    \label{eqn:fidelity_kernel}
\end{equation}
where $F(\rho,\sigma)$ is the Uhlmann fidelity~\cite{marian2012uhlmann} between two density matrices:
\begin{equation}
    F(\rho,\sigma)
    =
    \mathrm{Tr}
    \left[
    \sqrt{
    \sqrt{\rho}\sigma\sqrt{\rho}
    }
    \right].
    \label{eqn:uhlmann_fidelity}
\end{equation}
The predicted magnetic field for a test trajectory $\mathcal{R}$ is then given by
\begin{equation}
    \hat{B}(\mathcal{R})
    =
    \sum_{i=1}^{N}
    \alpha_i k(\mathcal{R},\mathcal{R}_i),
    \label{eqn:kernel_prediction}
\end{equation}
where the coefficients $\boldsymbol{\alpha}$ are obtained by solving
\begin{equation}
    (K+\lambda I)\boldsymbol{\alpha}=\mathbf{B}.
\end{equation}
Here, $K$ is the Gram matrix with entries $K_{ij}=k(\mathcal{R}_i,\mathcal{R}_j)$, and $\mathbf{B}$ contains the training magnetic-field values.

\subsection{Oracle and Tomography-Limited Quantum Kernels}

We evaluate two forms of the quantum kernel. The first is an oracle kernel computed from the exact pre-measurement density matrices $\rho_i(t_k)$. 
This setting is not intended to represent a directly available experimental measurement pipeline. 
Instead, it provides an upper bound on the performance that could be achieved if the learner had direct access to coherent quantum-state information before measurement-induced information loss.

The second is a tomography-limited kernel computed from reconstructed density matrices. 
In this setting, the density matrix at each evolution time is estimated from measured Pauli expectation values:
\begin{equation}
    \hat{\rho}(t_k)
    =
    \frac{1}{2}
    \left(
    I+
    \sum_{\alpha \in \{X,Y,Z\}}
    \braket{\sigma_{\alpha}(t_k)}
    \sigma_{\alpha}
    \right).
    \label{eqn:tomographic_reconstruction}
\end{equation}
The tomography-based reconstructed state is still constrained by finite-shot sampling, readout noise, and basis-measurement overhead.
Therefore, this setting provides a more realistic comparison between the quantum and classical learning methods.

The comparison between classical feature-based learning, tomography-limited quantum kernels, and oracle quantum kernels, therefore, isolates the central learning mechanism studied in this work. 
If the oracle quantum kernel outperforms the measurement-based models, the gain should signify the value of coherent quantum-state access.

\begin{figure}[b!]
    \centering
    \includegraphics[width=0.9\linewidth]{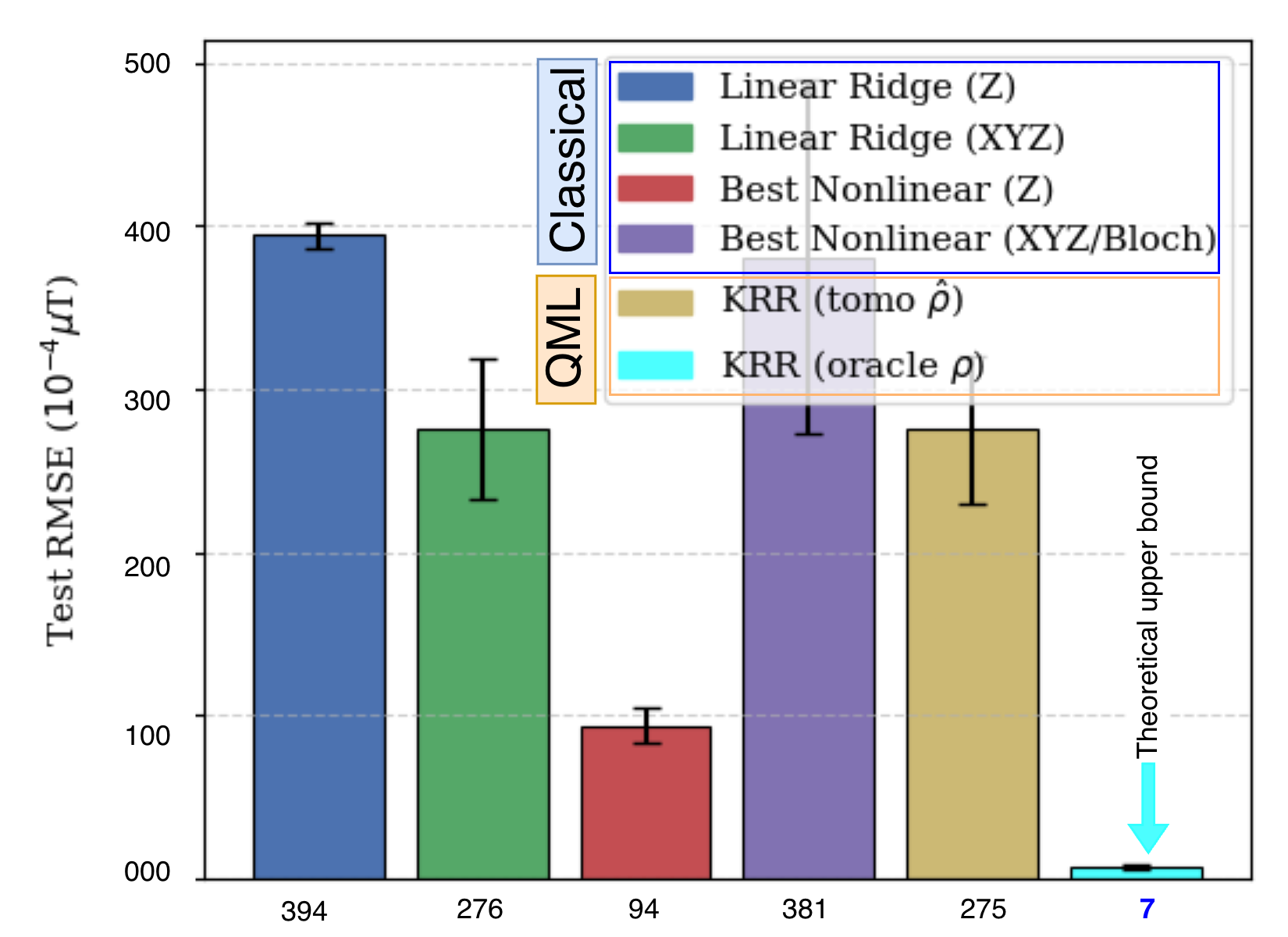}
    \caption{Test RMSE ($\mu$T) at 2048 shots (mean $\pm$ std): \footnotesize{The RMSE is significantly lower in the theoretical upper bound, which highlights its gap with the measurement-based learning models.}}
    \label{fig: res_2048}
\end{figure}

\section{Results}\label{sec: results}
We evaluated the magnetic field regression performance using 800 train and 100 test samples. For each sampled field value ($B_i\in[0.0, 2.0]\mu T$), Ramsey data is generated under a realistic dephasing noise model at multiple evolution times. We compared the performances of the following models:
\begin{enumerate}
    \item Linear ridge regression (LRR) on $Z$-only, as well as $XYZ$-measurement data.
    \item  Nonlinear classical models (RBF kernel ridge regression and multilayer perceptron (MLP)) using $XYZ$-measurement data.
    \item  Quantum kernel ridge regression (Q-KRR, using the Uhlmann fidelity~\cite{marian2012uhlmann}, equation~\ref{eqn:fidelity_kernel}, using (a) oracle density matrices, $\rho$, and (b) reconstructed density matrices using tomography, equation.\ref{eqn:tomographic_reconstruction}.
\end{enumerate}

Figure~\ref{fig: results} shows the test root-mean squared error (RMSE), produced by the models as a function of total number of shots (mean$\pm$std over seeds). We observed that the linear models show monotonic improvement, with XYZ features generally outperforming Z-only features at moderate-to-high shot counts. Whereas the nonlinear classical models (RBF/MLP) significantly outperform linear baselines across several regimes, particularly for Z-only features at higher shots, indicating that part of the estimation problem is nonlinear in the measurement statistics.

The oracle quantum kernel (KRR on exact $\rho$) provides a near-constant, substantially lower error floor ($\sim 7\times10^{-4} \mu T$), which represents a theoretical upper bound devoid of any loss due to measurement [Figure ~\ref{fig: res_2048}]. 
In contrast, the tomography-limited quantum kernel tracks the performance of strong classical XYZ-based models, which demonstrates that realistic measurement constraints significantly reduce the gap between the quantum and classical models.

These results indicate that quantum advantage in learning-based quantum sensing might primarily stem from access to coherent quantum-state information rather than from model expressivity alone.

\balance
\section{Conclusion}\label{sec: conclusion}
In this work, we investigated the role of data in quantum machine learning (QML) for quantum sensing via an NV center-inspired magnetometry simulation. 
We formulated magnetic field estimation as a supervised regression problem and compared classical learning models trained on post-measurement statistics with quantum kernel-based models trained on coherent pre-measurement quantum states. 
This comparison allowed us to isolate the effect of measurement-induced information loss and examine whether improved sensing performance arises from the learning model itself or from the type of data available to it.

Our results show that lossy measurements fundamentally constrain models trained solely on measured classical data.
While classical models can learn useful correlations from finite-shot measurement statistics, their performance is limited once quantum-state information has been collapsed into classical outcomes. 
In contrast, quantum models operating on coherent quantum-state representations achieve much lower estimation error.
They provide a theoretical upper bound on the sensing performance achievable when coherent quantum information is directly accessible for learning. 
These findings indicate that the potential advantage of QML in quantum sensing is not solely due to increased model complexity. Rather, it is strongly linked to the preservation and use of quantum coherence.

The study also highlights an important future direction for the design of quantum sensing systems. 
If sensing data are fully measured and converted into classical statistics before learning, a significant portion of the quantum advantage may already be lost. 
Therefore, future QML-enhanced sensing architectures should integrate sensing, state processing, and learning more closely within the quantum domain. 
Such quantum-native pipelines will reduce information loss and improve parameter estimation in near-term quantum sensing.
In future work, we will extend this analysis to hardware experiments, more complex noise models, and hybrid sensing-learning architectures that can operate under practical device limitations. 
Our experimental outcomes suggest that the path toward useful QML for quantum sensing depends not only on designing stronger learning models, but also on preserving coherent quantum information in the learning data.

\bibliographystyle{main}
\bibliography{main}

@article{rondin2014magnetometry,
  title={Magnetometry with nitrogen-vacancy defects in diamond},
  author={Rondin, Lo{\"\i}c and Tetienne, Jean-Philippe and Hingant, Thomas and Roch, Jean-Fran{\c{c}}ois and Maletinsky, Patrick and Jacques, Vincent},
  journal={Reports on progress in physics},
  volume={77},
  number={5},
  pages={056503},
  year={2014},
  publisher={IOP Publishing}
}

@article{degen2017quantum,
  title={Quantum sensing},
  author={Degen, Christian L and Reinhard, Friedemann and Cappellaro, Paola},
  journal={Reviews of modern physics},
  volume={89},
  number={3},
  pages={035002},
  year={2017},
  publisher={APS}
}

@article{schuld2019quantum,
  title={Quantum machine learning in feature Hilbert spaces},
  author={Schuld, Maria and Killoran, Nathan},
  journal={Physical review letters},
  volume={122},
  number={4},
  pages={040504},
  year={2019},
  publisher={APS}
}

@book{nielsen2010quantum,
  title={Quantum computation and quantum information},
  author={Nielsen, Michael A and Chuang, Isaac L},
  year={2010},
  publisher={Cambridge university press}
}

@article{hong2013nanoscale,
  title={Nanoscale magnetometry with NV centers in diamond},
  author={Hong, Sungkun and Grinolds, Michael S and Pham, Linh M and Le Sage, David and Luan, Lan and Walsworth, Ronald L and Yacoby, Amir},
  journal={MRS bulletin},
  volume={38},
  number={2},
  pages={155--161},
  year={2013},
  publisher={Cambridge University Press}
}

@article{schirhagl2014nitrogen,
  title={Nitrogen-vacancy centers in diamond: nanoscale sensors for physics and biology},
  author={Schirhagl, Romana and Chang, Kevin and Loretz, Michael and Degen, Christian L},
  journal={Annual review of physical chemistry},
  volume={65},
  number={1},
  pages={83--105},
  year={2014},
  publisher={Annual Reviews}
}

@article{rawat2026emerging,
  title={Emerging Paradigm of Quantum Machine Learning: Knowledge Insights and Future Prospects},
  author={Rawat, Keshav Singh and Sharma, Tarun},
  journal={IEEE Transactions on Artificial Intelligence},
  year={2026},
  publisher={IEEE}
}

@article{marian2012uhlmann,
  title={Uhlmann fidelity between two-mode Gaussian states},
  author={Marian, Paulina and Marian, Tudor A},
  journal={Physical Review A—Atomic, Molecular, and Optical Physics},
  volume={86},
  number={2},
  pages={022340},
  year={2012},
  publisher={APS}
}

@article{zeeman_split,
  title = {Zeeman Splitting of Nuclear Quadrupole Resonances},
  author = {Dean, C.},
  journal = {Phys. Rev.},
  volume = {96},
  issue = {4},
  pages = {1053--1059},
  numpages = {0},
  year = {1954},
  month = {Nov},
  publisher = {American Physical Society},
  doi = {10.1103/PhysRev.96.1053},
  url = {https://link.aps.org/doi/10.1103/PhysRev.96.1053}
}

@article{zhang2021distributed,
  title={Distributed quantum sensing},
  author={Zhang, Zheshen and Zhuang, Quntao},
  journal={Quantum Science \& Technology},
  volume={6},
  number={4},
  pages={043001},
  year={2021},
  publisher={IOP Publishing}
}


\end{document}